\documentclass[aps,pre,twocolumn,showpacs,groupedaddress,superscriptaddress,footinbib,longbibliography]{revtex4-2}

\usepackage{mathrsfs}
\usepackage{natbib,hyperref}
\setcitestyle{square,sort&compress,comma,numbers}
\hypersetup{colorlinks=true, citecolor=blue, urlcolor=blue, linkcolor=blue}
\usepackage[english]{babel}
\usepackage[utf8]{inputenc}
\usepackage{graphicx}
\usepackage[caption=false]{subfig}
\usepackage{amsmath}
\usepackage{amsfonts}
\usepackage{amssymb}
\usepackage{color,soul}
\setulcolor{red}
\usepackage{easyReview}
\usepackage{xcolor}
\usepackage{float}
\usepackage{pgfplots}
\usepgfplotslibrary{colormaps}
\pgfplotsset{
    colormap={mycolormap}{
        rgb255=(59,76,192)
        rgb255=(255,255,255)
        rgb255=(180,4,38)
    }
}

\usepackage[normalem]{ulem}
\newcommand\diff{\mathrm{d}}
\renewcommand{\vec}[1]{\boldsymbol{#1}}

\begin{document}

\title{First-passage-time statistics of active Brownian particles: A perturbative approach}
\author{Yanis Baouche}
\affiliation{Max-Planck-Institut f{\"u}r Physik komplexer Systeme, N{\"o}thnitzer Stra{\ss}e 38,
01187 Dresden, Germany}
\author{Magali Le Goff}
\affiliation{Institut f{\"u}r Theoretische Physik, Universit{\"a}t Innsbruck, Technikerstra{\ss}e 21A, A-6020 Innsbruck, Austria}
\author{Christina Kurzthaler}
\email{ckurzthaler@pks.mpg.de}
\affiliation{Max-Planck-Institut f{\"u}r Physik komplexer Systeme, N{\"o}thnitzer Stra{\ss}e 38,
01187 Dresden, Germany}
\affiliation{Center for Systems Biology Dresden, Pfotenhauerstra{\ss}e 108, 01307 Dresden, Germany}
\affiliation{Cluster of Excellence, Physics of Life, TU Dresden, Arnoldstra{\ss}e 18, 01062 Dresden, Germany}
\author{Thomas Franosch}
\email{thomas.franosch@uibk.ac.at}
\affiliation{Institut f{\"u}r Theoretische Physik, Universit{\"a}t Innsbruck, Technikerstra{\ss}e 21A, A-6020 Innsbruck, Austria}

\begin{abstract} 
We study the first-passage-time (FPT) properties of active Brownian particles to reach an absorbing wall in two dimensions. Employing a perturbation approach we obtain exact analytical predictions for the survival and FPT distributions for small P{\'e}clet numbers, measuring the importance of self-propulsion relative to diffusion. While randomly oriented active agents reach the wall faster than their passive counterpart, their initial orientation plays a crucial role in the FPT statistics.  Using the median as a metric, we quantify this anisotropy and find that it becomes more pronounced at distances where persistent active motion starts to dominate diffusion. 
\end{abstract} 

\maketitle

\section{Introduction}

The first-passage-time (FPT) statistics appear as a natural way of quantifying the efficiency of a variety processes. 
From catalytic reactions in chemistry~\cite{chmeliovFluorescenceBlinkingSingle2013, robinSinglemoleculeTheoryEnzymatic2018}
to economy where call/puts options are subject to time windows~\cite{wilmottPaulWilmottIntroduces2010, sazukaDistributionFirstpassageTimes2009}, its applications to physical sciences are numerous~\cite{rednerGuideFirstpassageProcesses2001}. FPT statistics are also of paramount importance in biological systems, in particular for swimming microorganisms~\cite{polizziMeanFirstPassageTimes2016, otteStatisticsPathogenicBacteria2021, suarezSpermTransportFemale2006, godecFirstPassageTime2016,caraglioLearningHowFind2024}, such as bacteria, protozoa or algae. Their self-propulsion mechanisms are the result of a long evolution and adaptation to their surroundings to both optimize their survival strategies, through efficient foraging and escaping from harm, and achieve biological functions~\cite{laugaHydrodynamicsSwimmingMicroorganisms2009, bechingerActiveParticlesComplex2016, laugaFluidDynamicsCell2020, kurzthalerOutofequilibriumSoftMatter2023}. Despite its importance for microbiology and the design of efficient nano-technological applications~\cite{erkocMobileMicrorobotsActive2019,alapanMicroroboticsMicroorganismsBiohybrid2019, volpeRoadmapAnimateMatter2024}, such as targeted drug delivery systems, analytical predictions quantifying the time it takes active agents to achieve certain tasks remain sparse~\cite{angelaniFirstpassageTimeRuntumble2014, malakarSteadyStateRelaxation2018,basuActiveBrownianMotion2018}.

A large artillery of methods has been developed and studied to address FPT statistics in various contexts.  
They usually involve solving a backward Fokker-Planck equation, the renewal equation or continuous-time random walk frameworks~\cite{masoliverRandomProcessesFirstpassage2018, kampenStochasticProcessesPhysics2007}, the Feynman-Kac correspondence~\cite{oksendalStochasticDifferentialEquations2007} or  martingale theory~\cite{baldiMartingalesMarkovChains2002,lawlerIntroductionStochasticProcesses2006}. 
The vast majority of physical models typically builds on the Wiener process of standard diffusion, which is well known for having a normalizable FPT probability density, meaning that the system inevitably reaches its target, but also for having a divergent mean FPT~\cite{rednerGuideFirstpassageProcesses2001, balakrishnanFirstPassageTime2021}. 
Extensions include the Heston and Feller models with additional degrees of freedom~\cite{masoliverEscapeProblemStochastic2008, masoliverRandomProcessesFirstpassage2018}, such as external forcing or multiplicative noise, providing analytical predictions for economics and neurosciences. Furthermore, supplementing the standard diffusion model with another source of stochasticity, the paradigmatic example being stochastic resetting~\cite{evansDiffusionStochasticResetting2011c,masoliverTelegraphicProcessesStochastic2019,evansStochasticResettingApplications2020a}, where a procedure is randomly reset to a certain state, may enable a finite mean FPT. Yet, the signature of this divergence still manifests itself in several processes and thus other metrics, such as the median or the mode, have been brought forward~\cite{belanMedianModeFirst2020}.

Deviations from standard diffusion are expected to become particularly important for active transport phenomena. 
Paradigmatic models to describe the dynamics of these out-of-equilibrium agents include the run-and-tumble particle, describing the motion of bacteria, algae or other flagellated organisms \cite{bergChemotaxisEscherichiaColi1972, bergChemotaxisBacteriaGlass1990,woolleyMotilitySpermatozoaSurfaces2003,codlingRandomWalkModels2008,catesDiffusiveTransportDetailed2012,zhaoQuantitativeCharacterizationRunandtumble2024a,kurzthalerCharacterizationControlRunandTumble2024a}, and the active Brownian (ABP) particle~\cite{howseSelfMotileColloidalParticles2007a, hagenBrownianMotionSelfpropelled2011, romanczukActiveBrownianParticles2012a, sevillaTheoryDiffusionActive2014,kurzthalerIntermediateScatteringFunction2016, kurzthalerProbingSpatiotemporalDynamics2018a}, capturing the effects emerging from rotational diffusion. While the FPT statistics of run-and-tumble agents have been studied in one dimension \cite{angelaniFirstpassageTimeRuntumble2014, malakarSteadyStateRelaxation2018}, less is known for ABPs. Recent work has provided analytical predictions for the FPT probability density of an ABP escaping a circular boundary~\cite{ditrapaniActiveBrownianParticles2023b}, revealing interesting features due to the interplay of activity and confinement. Insights for an ABP reaching an absorbing wall in two dimensions based on simulations and scaling arguments have been obtained in the limit of infinite activity~\cite{basuActiveBrownianMotion2018}, where the survival distribution exhibits an anomalous slow decay, $S(t) \propto t^{-1/4}$ for $t \to \infty$,  for agents close and initially oriented parallel to the boundary, in contrast to their passive counterpart, $S(t) \propto t^{-1/2}$. However, in general, analytical predictions of the FPT distributions and (if existent) the mean or median FPT, required to establish a fundamental understanding of the underlying physical processes, however, are lacking.

Here, we study the FPT statistics of an ABP to reach an absorbing wall in two dimensions using a perturbation approach for small activity.
To zeroth order we recover the Green's function of a passive agent in the presence of the wall, allowing us to successively solve each order of our perturbation theory analytically. We then discuss the influence of the initial angle and initial position on the FPT and survival distributions. We finally highlight the anisotropy of the FPT statistics, measuring the dependence of the median FPT on the initial orientation, and find that, at moderate P{\'e}clet numbers, it becomes maximal when active motion dominates over diffusion.

\section{Theory \label{sec:theory}} 
\begin{figure}[tp]
\includegraphics[width = \linewidth]
{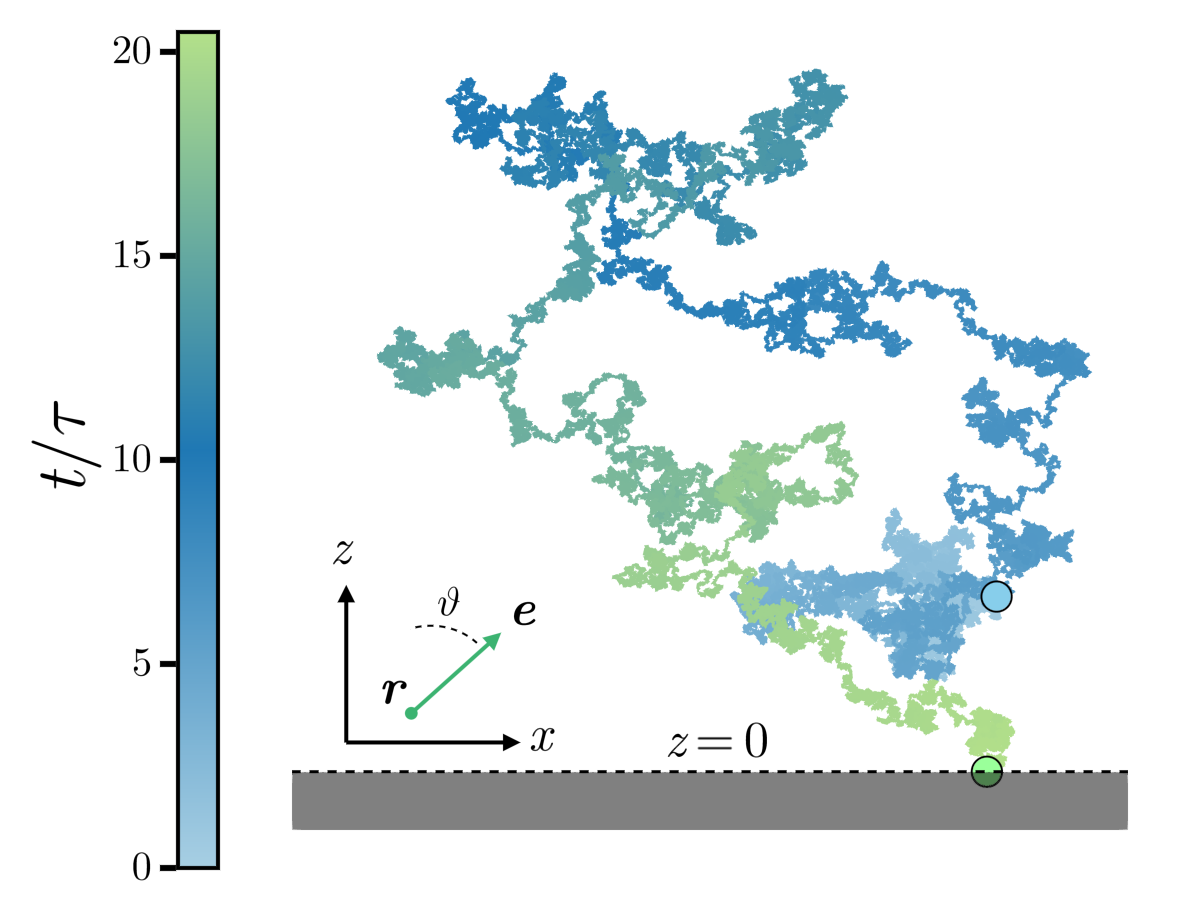}
\caption{Trajectory of an ABP with an absorbing boundary at $z=0$. Circles indicate the beginning and final point of the trajectory. Different colors correspond to different times $t/\tau$. ({\it Inset}) Sketch of the position $\vec{r}$ and orientation $\vartheta$, measured relative to the vertical direction ($\vec{e}_z$), of the active particle. \label{fig:schematic}}
\end{figure}
We consider an ABP in the two-dimensional plane $(O, x, z)$ moving at constant speed $v$ along its instantaneous orientation $\vec{e}(\vartheta(t)) =(\sin(\vartheta(t)), \cos(\vartheta(t)))$, where $\vartheta(t)$ denotes the polar angle measured from $\vec{e}_z$ such that for $\vartheta(t)=0$ the particle faces against the wall [Fig.~\ref{fig:schematic}({\it inset})].  The agent is subject to translational and rotational diffusion with diffusion coefficients $D$ and $D_{\mathrm{rot}}$, respectively. The equations of motion for the position $\boldsymbol{r} =(x,z)$ and orientation $\vartheta$ read
\begin{subequations}
\begin{align}
\frac{\diff \vec{r}}{\diff t} &= v\vec{e} +\sqrt{2D}\vec{\eta}, \label{eq:stoch_r}\\
\frac{\diff \vartheta}{\diff t} &= \sqrt{2D_{\mathrm{rot}}}\xi \label{eq:stoch_theta},
\end{align}
\end{subequations}
where $\vec{\eta}(t)$ and $\xi(t)$ are independent Gaussian white noises of zero mean and delta correlated variance, $\langle \eta_{i}(t) \eta_{j}(t') \rangle = \delta_{ij} \delta(t-t')$ for $i,j =1,2$ and $\langle \xi(t) \xi(t') \rangle = \delta(t-t')$.
The associated probability density of a particle to be at~$\vec{r}$ with orientation~$\vartheta$ at time $t$ having started at~$\vec{r}_0$ with orientation~$\vartheta_0$ at $t=0$ is denoted by $\mathbb{P}(\vec{r}, \vartheta, t|\vec{r}_0, \vartheta_0)$. It obeys the Fokker-Planck equation with absorbing boundary condition at the wall ($z=0$) 
\begin{subequations}
    \begin{align}
        \partial_{t} \mathbb{P} = - v \vec{e} \cdot \boldsymbol{\nabla} \mathbb{P} + D_{\mathrm{rot}} \partial_{\vartheta}^{2} \mathbb{P} + D \nabla^{2} \mathbb{P},\\
        \mathbb{P}(\vec{r}, \vartheta,t=0 | \vec{r}_{0}, \vartheta_{0}) = \delta(\vec{r}-\vec{r}_{0}) \delta(\vartheta - \vartheta_{0}),\\
        \mathbb{P}(x,z=0, \vartheta,t| \boldsymbol{r}_{0},\vartheta_{0})  = 0 \quad  \forall t \in \mathbb{R}^+. \label{eq:BC}
    \end{align} 
\end{subequations}
We further rescale the equations of motion with the particle's hydrodynamic radius $a$ as characteristic length scale, $\vec{r}=a\vec{R}$, and the diffusive time $\tau = a^2/D$ as characteristic time scale, $t= \tau T$. Integrating over the $X-$direction (corresponding to the direction parallel to the wall), we arrive at the non-dimensional equation for $\mathbb{P}(Z,\vartheta, T|Z_{0},\vartheta_{0})$:  
\begin{align}
    \partial_{{T}} \mathbb{P} = - {\mathrm{Pe}} \cos(\vartheta) \partial_{Z} \mathbb{P} +  \partial_{{Z}}^2\mathbb{P} +\gamma \partial_{\vartheta}^{2}  \mathbb{P}. \label{eq:FP}
    \end{align}
Here, $\mathrm{Pe}=va/D$ denotes the P{\'e}clet number, quantifying the relative importance of active motion and diffusion. Using the Stokes-Einstein-Sutherland relation for the hydrodynamic radius $a$, we have $\gamma=D_{\mathrm{rot}} a^2/D = 3/4$.

We are interested in studying the FPT statistics of an ABP to reach the wall, having started at initial distance $Z_{0}>0$ from the wall with initial orientation $\vartheta_{0}$. Therefore, we introduce the survival probability $S(T|Z_0, \vartheta_0)$, which measures the probability that a particle has not reached the wall up to time $T$. It is obtained by marginalizing over $Z$ 
\begin{align}
{S}({T} | {Z}_{0}, \vartheta_{0} ) &= \int_{0}^{\infty} \mathrm{d}Z ~ \mathbb{P}(Z ,T | Z_{0}, \vartheta_{0}),
\end{align}
and provides direct access to the FPT probability density,  measuring the distribution of the first times at which the particle arrives at the wall (or equivalently the rate of loss of particles in the half-space): 
\begin{align}
F({T} | {Z}_{0}, \vartheta_{0}) = - \frac{\diff }{\diff {T}}S({T} | {Z}_{0}, \vartheta_{0}).
\end{align}
To derive analytical expressions of these quantities, we solve Eq.~\eqref{eq:FP} by adopting a  perturbation scheme in the P{\'e}clet number
\begin{equation}
        \mathbb{P} = \mathbb{P}_{0} + \mathrm{Pe} \ \mathbb{P}_{1} + \mathrm{Pe}^{2} \ \mathbb{P}_{2} + O(\mathrm{Pe}^3), \label{eq:expansion}
\end{equation}
    where $\mathbb{P}_0$ corresponds to the propagator of a passive particle near an absorbing wall, and $\mathbb{P}_1$ and $\mathbb{P}_2$ represent the first and second-order perturbations, respectively. In what follows, we will present an iterative scheme to compute those analytically. 

First, we transform to Laplace space $T \mapsto s$: 
\begin{align}
\widehat{\mathbb{P}}(Z, \vartheta, s|Z_0, \vartheta_0) = \int_0^\infty \diff T\ e^{-sT}\ \mathbb{P}(Z, \vartheta, T|Z_0, \vartheta_0).
\end{align}
Taking the Laplace transform of Eq.~\eqref{eq:FP} leads to 
    \begin{equation}
        (s-\mathcal{H})\widehat{\mathbb{P}} = \delta(Z-Z_{0})\delta(\vartheta-\vartheta_{0}), \label{eq:Laplace}
    \end{equation}
where the operator $\mathcal{H}$ comprises the unperturbed operator $\mathcal{H}_0$ and the perturbation $\mathcal{V}$:
\begin{subequations}
    \begin{align}
        \mathcal{H}&= \partial_{Z}^{2}  +\gamma \partial_{\vartheta}^{2} +{\mathrm{Pe}}  (-\cos(\vartheta) \partial_{Z})\\
        &=: \mathcal{H}_{0} + \mathrm{Pe}\mathcal{V}.
    \end{align}
\end{subequations}
Inserting the Laplace transform of the expansion [Eq.~\eqref{eq:expansion}] into Eq.~\eqref{eq:Laplace}, leads to a set of coupled equations for the perturbations
\begin{subequations}
    \begin{align}
        (s-\mathcal{H}_{0}) \widehat{\mathbb{P}}_{0} &=\delta(Z-Z_{0})\delta(\vartheta-\vartheta_{0}), \label{eq:G}\\
        (s-\mathcal{H}_{0}) \widehat{\mathbb{P}}_{1} &= \mathcal{V} \widehat{\mathbb{P}}_{0} \label{perturb_eq},\\
    (s-\mathcal{H}_{0}) \widehat{\mathbb{P}}_{2} &= \mathcal{V} \widehat{\mathbb{P}}_{1}, \label{perturb_eq2}
    \end{align}
\end{subequations}
which can be extended to arbitrary order in $\mathrm{Pe}$. The structure of the equations suggests an iterative solution strategy for each order of the following form. 
The solution $\widehat{\mathbb{P}}_{i+1}$ of order $i+1$ can be computed using the Green's function $G$ and the $i$-th order solution~$\widehat{\mathbb{P}}_{i}$:
\begin{align}
    &\widehat{\mathbb{P}}_{i+1}(Z,\vartheta,s | Z_{0}, \vartheta_{0}) =\label{eq:pi}\\ &\int_0^\infty \mathrm{d}Z'  \int_{0}^{2\pi} \mathrm{d}\vartheta' ~ G(s,Z,\vartheta ,  Z', \vartheta')~ [\mathcal{V} \widehat{\mathbb{P}}_{i}]( Z', \vartheta',s | Z_{0},\vartheta_{0}), \notag
\end{align}
where $G(\cdot)$ denotes the Green's function solving 
\begin{equation}
    (s-\mathcal{H}_{0})G=\delta(Z-Z')\delta(\vartheta-\vartheta' ),
    \label{eq:greens_func}
\end{equation}
with boundary condition~$G(s, Z=0,\vartheta, Z', \vartheta')=0$. We note that the zeroth-order solution coincides with the Green's function,  $\widehat{\mathbb{P}}_{0}(Z,\vartheta,s | Z_{0}, \vartheta_{0})=G(s,Z,\vartheta, Z_{0}, \vartheta_{0})$, and thus knowledge of $G$ is the fundamental ingredient for computing higher-order contributions. 

\subsection{Green's function}
To compute the Green's function we employ an image method, akin to electrodynamics. In particular, we first compute the Green's function for an unbounded domain~$G^{u}$, ignoring the absorbing boundary condition. As starting point, we decompose it in terms of the angular modes
\begin{equation}
    G^{u}(s,Z,\vartheta, Z', \vartheta') = \frac{1}{2\pi} \sum_{\ell=-\infty}^\infty f_{\ell}(s,Z, Z', \vartheta') e^{-i \ell \vartheta}, \label{eq:green_unbounded_modes}
\end{equation}
with coefficients
\begin{equation}
    f_{\ell}(s,Z, Z', \vartheta') = \int_{0}^{2\pi} \mathrm{d}\vartheta \  G^{u}(s,Z,\vartheta, Z', \vartheta') e^{i\ell \vartheta}.
\end{equation}
Next, we insert the expansion [Eq.\eqref{eq:green_unbounded_modes}] into Eq.~\eqref{eq:greens_func} and multiply it with $e^{i\ell \vartheta}$. Integrating over the final orientation $\vartheta$ yields
\begin{equation}
    ( p_{\ell}^{2} -\partial_{Z}^{2}) f_{\ell}(s,Z, Z', \vartheta') = \delta(Z-Z')e^{i\ell \vartheta'},
\end{equation}
with the abbreviation~$p_{\ell}^{2} =s + \gamma \ell^{2}$. Further progress can be made by taking the spatial Fourier transform $Z \mapsto K$, 
$\widetilde{f}_{\ell}(s,K, Z', \vartheta') = \int_{\mathbb{R}} \mathrm{d}Z \ f_{\ell}(s,Z, Z', \vartheta') e^{-iKZ}$. We thus arrive at 
\begin{equation}
    (p_{\ell}^{2} + K^{2}) \widetilde{f}_{\ell}(s,K, Z', \vartheta')=e^{-iKZ'}e^{i\ell \vartheta'}.
\end{equation}
To perform the inverse transform, we use the residue theorem:
\begin{equation}
    f_{\ell}(s,Z, Z', \vartheta') = 
        \frac{1}{2p_{\ell}}e^{-p_{\ell}|Z-Z'|} e^{i \ell \vartheta'} 
\end{equation}
Thus the solution for the Green's function in an unbounded domain reads
\begin{equation}
    G^{u}(s,Z, \vartheta, Z', \vartheta') =  \frac{1}{2\pi}\sum_{\ell=-\infty}^{\infty} \frac{1}{ 2p_{\ell}}e^{-p_{\ell}|Z-Z'|} e^{i \ell (\vartheta' -\vartheta)}.
\end{equation}

To enforce the presence of the wall at $Z=0$ in terms of the absorbing boundary condition, we construct an image solution of the form $G^{*}= G^{u}(s,Z, \vartheta, -Z_0, \vartheta')$, corresponding to the Green's function of a diffusing particle with initial position at the opposite side of the wall $Z_0\to-Z_0$. The full solution can then be obtained by $G = G^{u} -G^*$, leading to 
\begin{align}
\begin{split}
           G(s,Z, \vartheta,Z', \vartheta') &= \frac{1}{2\pi} \sum_{\ell=-\infty}^{\infty}\frac{1}{2 p_{\ell}}e^{i \ell (\vartheta' -\vartheta)}\\ 
        &\times\left( e^{-p_{\ell}|Z-Z'|} -e^{-p_{\ell}|Z+Z'|}\right),  
\end{split}
\end{align}
which serves as building block for computing the propagator [Eq.~\eqref{eq:expansion}].

\subsection{Zeroth-order solution: Brownian particle}
We recall that the zeroth-order solution is given by the Green's function: $\widehat{\mathbb{P}}_{0}(Z,\vartheta,s | Z_{0}, \vartheta_{0})=G(s,Z,\vartheta, Z_{0}, \vartheta_{0})$. To compute the survival probability and the FPT distribution of the zeroth-order solution, we first marginalize over the final orientation $\vartheta$, yielding:
\begin{subequations}
\begin{align}
           &\widehat{\mathbb{P}}_0(s,Z|Z_{0}, \vartheta_{0}) =  \int_0^{2\pi} \ \diff \vartheta \ G(s,Z, \vartheta, Z_{0}, \vartheta_{0}),\\
        &= \frac{1}{2\sqrt{s}} \left( e^{-\sqrt{s}|Z-Z_{0}|} -e^{-\sqrt{s}|Z+Z_{0}|}\right).
\end{align}
\end{subequations}
We note that the dependence on the initial orientation $\vartheta_0$ drops out as translation and rotation are uncoupled.
In Laplace space, the zeroth-order survival probability is obtained via marginalizing over $Z$ 
\begin{subequations}
\begin{align}
\widehat{S}_0(s | Z_{0}) &= \int_{0}^{\infty} \mathrm{d}Z ~  \widehat{\mathbb{P}}_0(s,Z|Z_{0}, \vartheta_{0}),\\
&= \frac{1}{s}\left(1-e^{-\sqrt{s}Z_{0}} \right),
\label{theory:perturb_s0}
\end{align}
\end{subequations}
and  provides immediate access to the FPT probability density via 
\begin{align}
\widehat{F}_0(s | Z_{0}) = 1-s\widehat{S}_{0}(s | Z_{0}) = e^{-\sqrt{s}Z_0}.
\end{align}
Going back to the temporal domain, we recover the well-established result for the FPT probability density of a passive Brownian tracer (in non-dimensional form)~\cite{rednerGuideFirstpassageProcesses2001}:
\begin{align}
F_0(T | Z_{0}) =  \frac{Z_{0}}{\sqrt{4 \pi T^3}}  e^{-\frac{Z_{0}^2}{4T}},
\end{align}
representing the probability density function of a L\'{e}vy distribution with asymptotic scaling~$F_{0} \propto  T^{-3/2}$ for ~$T \to \infty$.

\subsection{First-order perturbation}
Inserting $\widehat{\mathbb{P}}_{0}$ and $G$ into Eq.~\eqref{eq:pi} provides analytical expressions for $\widehat{\mathbb{P}}_{1}$, which, due to the presence of absolute values, is split into two parts:
\begin{subequations}
\begin{align}
    &\widehat{\mathbb{P}}_{1}^{-}(s, Z |Z_{0}, \vartheta_{0}) = \frac{\cos(\vartheta_{0})}{2 \gamma}  
    \Bigg[ e^{-(Z+Z_{0})\sqrt{s} }  -2e^{-Z_{0} \sqrt{s+\gamma} - Z \sqrt{s}} \notag \\
    &-e^{(Z-Z_{0})\sqrt{s}} + e^{-(Z + Z_{0}) \sqrt{s+\gamma}} + e^{(Z - Z_{0}) \sqrt{s+\gamma}}  \Bigg] ~ \mathrm{for} \ Z \leq Z_{0},\\ 
    &\widehat{\mathbb{P}}_{1}^{+}(s, Z |Z_{0}, \vartheta_{0}) = \frac{\cos(\vartheta_{0})}{2 \gamma}  
    \Bigg[ e^{-(Z+Z_{0})\sqrt{s} }  -2e^{-Z_{0} \sqrt{s+\gamma} - Z \sqrt{s}} \notag \\
    &+e^{(Z_{0}-Z)\sqrt{s}} + e^{-(Z + Z_{0}) \sqrt{s+\gamma}} - e^{(Z_{0} - Z) \sqrt{s+\gamma}}  \Bigg] ~ \mathrm{for} \ Z \geq Z_{0}.
\end{align}
\end{subequations}
The first-order correction to the survival probability is obtained by marginalizing over $Z$
\begin{align}
    \widehat{S}_{1}(s | Z_{0}, \vartheta_{0} ) &=
    \int_{0}^{\infty} \mathrm{d}Z ~ \widehat{\mathbb{P}}_{1}(s, Z |Z_{0}, \vartheta_{0}), 
    \notag \\
    &= \frac{\cos(\vartheta_{0})}{\gamma} \left( \frac{e^{-\sqrt{s}Z_{0} }}{\sqrt{s}}  - \frac{e^{-\sqrt{s+\gamma}Z_{0}}}{\sqrt{s}}  \right),
    \label{eq:perturb_s1}  
\end{align}
where we have split the integration domain $[0,\infty) = [0, Z_{0}] \cup [Z_{0}, \infty)$ to integrate over the piece-wise function $\widehat{\mathbb{P}}_{1}$.  The first-order correction to the FPT distribution can then be directly computed in Laplace space:
\begin{align}
    \widehat{F}_{1}&(s | Z_{0}, \vartheta_{0} ) =-s\widehat{S}_1(s|Z_0, \vartheta_0), \notag \\ 
    &=  \frac{\sqrt{s}}{\gamma}(e^{-\sqrt{s+\gamma}Z_{0}}-e^{-\sqrt{s}Z_{0} } ) \cos(\vartheta_{0}).
        \label{theory:perturb_f1}
\end{align}

Interestingly, the first term in Eq.~\eqref{eq:perturb_s1} possesses an analytical inverse Laplace transform $\mathcal{L}^{-1} : s \mapsto T$ of the form
\begin{align} 
\mathcal{L}^{-1} \left\{ 
 \frac{e^{-\sqrt{s}Z_{0}}}{\sqrt{s}} \right\} &=  \frac{e^{- \frac{ Z_{0}^{2}}{4T} }}{\sqrt{\pi T }},
\end{align}
while the second term $e^{-\sqrt{s+\gamma}}/\sqrt{s}$ carries the mark of the rotational diffusion $\gamma$ in the exponential but not in the prefactor $1/\sqrt{s}$, thus making the inverse Laplace transform challenging. However, one can still make further analytical progress. Following Ref.~\cite{puriInverseLaplaceTransforms1988}, the inverse transform of the second term is expressible as a parametric integral:
\begin{align}
    \mathcal{L}^{-1} \left\{ \frac{e^{-Z_{0} \sqrt{s+\gamma}} }{\sqrt{s}} \right\} &= \frac{1}{\pi} \Bigg[ \int_{0}^{\infty} \diff x \frac{ \cos \left(Z_{0} \sqrt{x} \right)}{\sqrt{x+\gamma}} e^{-(x+\gamma)T} \notag \\
    &+ \int_{0}^{\gamma} \diff x \frac{e^{-Z_{0} \sqrt{\gamma-x} - xT }}{\sqrt{x}}  \Bigg].
\end{align}
Collecting these results, the full first-order perturbation of the survival probability becomes:
\begin{align}
&S_{1}(T | Z_{0}, \vartheta_{0}) = 
\frac{\cos(\vartheta_{0})}{\gamma} \Bigg[ 
  -\frac{1}{\pi}  \int_{0}^{\infty}\diff x  \frac{ \cos \left(Z_{0} \sqrt{x} \right)}{\sqrt{x+\gamma}} e^{-(x+\gamma)T}  \notag \\  
&- \frac{1}{\pi}\int_{0}^{\gamma}\diff x  \frac{e^{-Z_{0} \sqrt{\gamma-x} - xT }}{\sqrt{x}}  + \frac{e^{- \frac{ Z_{0}^{2}}{4T} }}{\sqrt{\pi T }}\Bigg].
\end{align}
Taking the negative time derivative, we obtain the FPT probability density
\begin{align}
&F_{1}(T | Z_{0}, \vartheta_{0}) = \\  
&-\frac{\cos(\vartheta_{0})}{\gamma} \Bigg[ 
\frac{1}{\pi}\int_{0}^{\infty}\diff x  \cos(Z_{0} \sqrt{x}) \sqrt{x+\gamma} e^{-(x+\gamma)T}  \notag\\ 
&+ \frac{1}{\pi} \int_{0}^{\gamma} \diff x \sqrt{x} e^{-Z_{0} \sqrt{\gamma-x} - xT }+ \frac{e^{- \frac{ Z_{0}^{2}}{4T} }}{\sqrt{16\pi T^{5}}}(Z_{0}^{2}-2T) \Bigg]. \notag
\end{align}
Even though the presence of the $\gamma$-term prevents the existence of an antiderivative, the previous expression allows computing the first-order correction at a lighter cost than performing an inverse Laplace transform.
The above equations contrast with the exponentially-damped power laws of Brownian particles, highlighting the role of activity and varying swimming angle through $\gamma$.

\subsection{Second-order perturbation}
To compute the next order, we insert $\widehat{\mathbb{P}}_{1}$ and $G$ into Eq.~\eqref{eq:pi} and follow the same procedure as before to obtain the second-order correction to the survival probability
\begin{equation}
    \begin{split}
        &\widehat{S}_{2}(s,| Z_{0}, \vartheta_{0}) = - \frac{e^{-Z_{0} (\sqrt{s} + \sqrt{s+\gamma} + \sqrt{s+4\gamma})}}{24 \sqrt{s} \sqrt{s+\gamma} \gamma^{2}} \Big[6 e^{Z_{0}\sqrt{s+4\gamma}} \times \\
        &\left( 2e^{\sqrt{s}Z_{0}}(s+\gamma) + e^{Z_{0}\sqrt{s+\gamma}} (-2s-2\gamma + Z_{0} \gamma \sqrt{s+\gamma}) \right) + \\ 
        & \Big( 3e^{Z_{0} (\sqrt{s+\gamma} + \sqrt{s+4\gamma})}\sqrt{s} \sqrt{s+\gamma} - 4e^{Z_{0}(\sqrt{s} + \sqrt{s+4\gamma})}(s+\gamma) \\
        &+ e^{Z_{0}(\sqrt{s}+\sqrt{s+\gamma})} (4s+4\gamma -3\sqrt{s}\sqrt{s+\gamma})  \Big) \cos(2\vartheta_{0}) \Big].  
        \label{eq:perturb_s2}
    \end{split}
\end{equation}
In the same fashion as before, we compute the second-order correction to the FPT distribution by
\begin{equation}
    \widehat{F}_{2}(s | Z_{0}, \vartheta_{0} ) = -s\widehat{S}_{2}(s | Z_{0}, \vartheta_{0}).
    \label{theory:perturb_f2}
\end{equation}
As an analytical inverse Laplace transform renders difficult, we compute its inverse numerically. Higher-order perturbations can be readily obtained through a recursive scheme, which we outline in Appendix~\ref{appendix:recursive_scheme}.
\begin{figure*}[ht]
\centering
  \includegraphics[width=\textwidth]{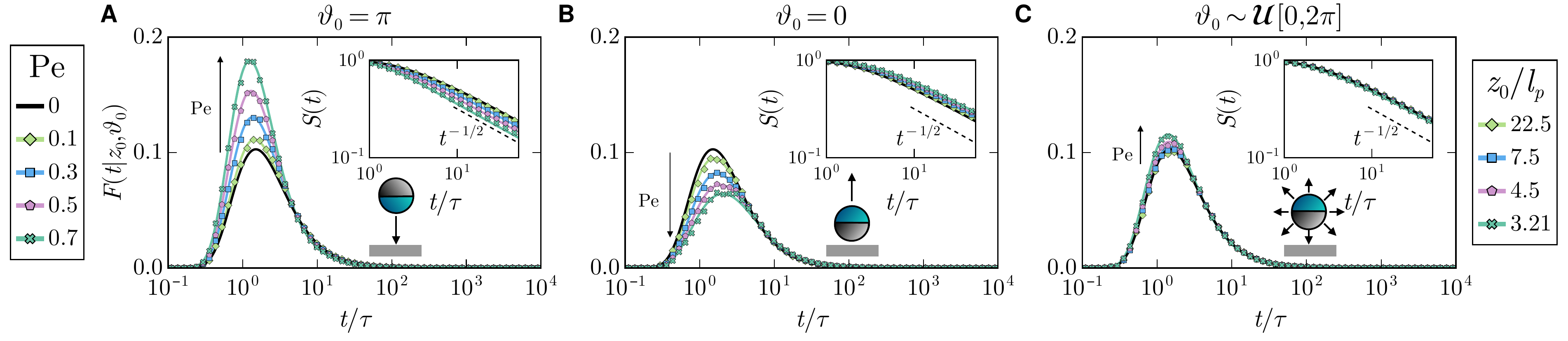}
     \caption{FPT probability density and survival probability (insets) as a function of time $t$ for three different initial angles~$\vartheta_{0}$: the particle is initially \textbf{A}~facing the wall, \textbf{B}~facing against the wall, and \textbf{C}~randomly oriented with an angle drawn from a uniform distribution $\mathcal{U}[0, 2\pi]$. Here, the initial position is $z_{0}/a=3$. Solid lines and symbols denote theory and simulations for different P{\'e}clet numbers, respectively.}
    \label{fig:fpt_survival}
\end{figure*}
\begin{figure*}[ht]
\includegraphics[width = \linewidth]{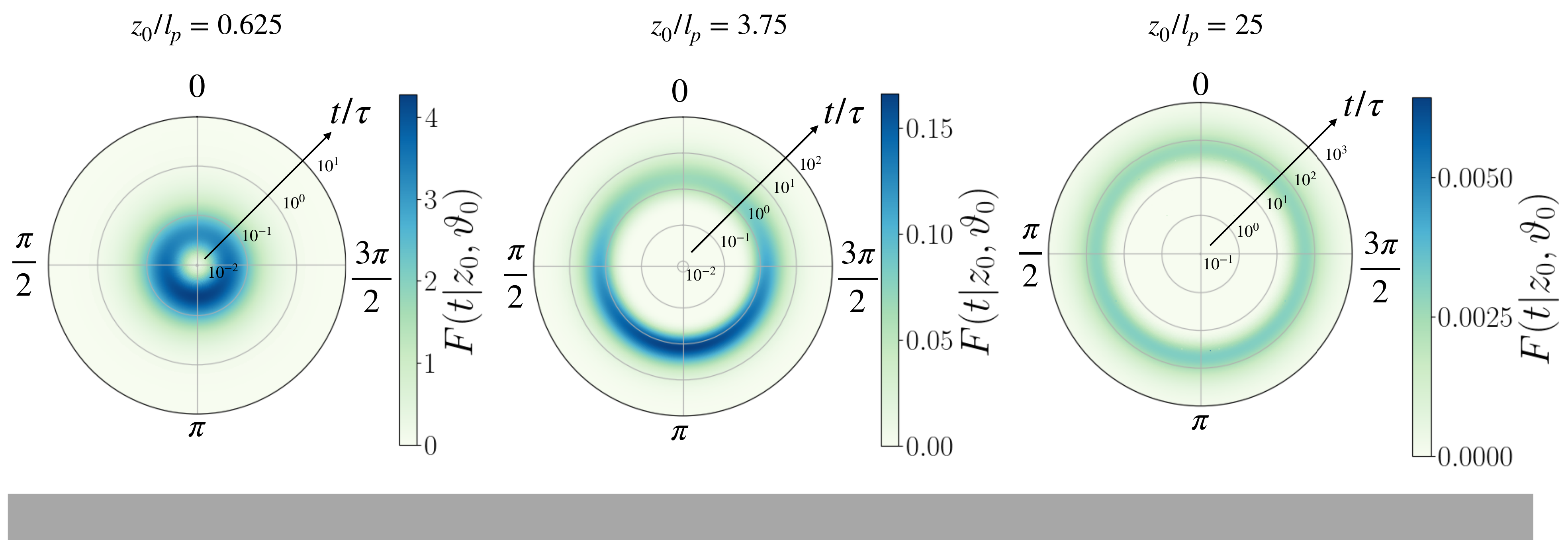}
\caption{Heatmap of the FPT probability density $F(t|z_{0}, \vartheta_{0})$ for different initial angles $\vartheta_0$ and initial positions $z_{0}$, rescaled by the persistence length $l_p$. The radial direction indicates the time $t$ and the angular direction indicates $\vartheta_{0}$. Here, the P{\'e}clet number is $\mathrm{Pe}=0.6$. }
\label{fig:fpt_angle}
\end{figure*}

\section{Results} 
We discuss our results for the FPT probability densities of ABPs with respect to their initial orientation $\vartheta_0$ and position $z_0$ in Sec.~\ref{sec:results_FPTdist} and comment on the contributions of the different orders of our perturbation theory in Sec.~\ref{sec:results_FPTorders}. We further quantify these findings in terms of the median in Sec.~\ref{sec:results_median} and provide a measure for the anisotropic nature of the FPT statistics in Sec.~\ref{sec:results_anisotropy}.

\subsection{First-passage time distributions \label{sec:results_FPTdist}}
We begin our analysis with the FPT probability density, where, starting from the perturbations in Eqs. \eqref{theory:perturb_f1}-\eqref{theory:perturb_f2}, we compute $F$ (resp. $S$) up to second order in the P{\'e}clet number. Our results are shown in Fig.~\ref{fig:fpt_survival} for the FPT probability density, as well as the survival probability (inset), for several initial orientations $\vartheta_{0}$ and P{\'e}clet numbers $\mathrm{Pe}$. Overall, the results match the intuition, in the sense that a particle initially oriented towards the wall (Fig.~\ref{fig:fpt_survival}~\textbf{A}) displays a distribution that is more peaked at shorter times (in comparison to the Brownian case) and that has less weight in its forward tail, as $\mathrm{Pe}$ increases. The survival probability respectively decreases with activity, as the particle is faster at reaching the wall. Conversely, an agent that is initially facing away from the wall (Fig. \ref{fig:fpt_survival}~\textbf{B}) will reach it at later times and has a higher chance of surviving in the half-space for a longer time.

We further investigate the pure effect of activity by averaging over the initial orientation. Integrating $F$ (resp. $S$) over $\vartheta_{0}$ results in a vanishing first-order perturbation [Eqs.~\eqref{eq:perturb_s1},\eqref{theory:perturb_f1}] and a major part of the second-order perturbation [Eq.~\eqref{eq:perturb_s2}]. In fact, this procedure leads to the vanishing of all odd-order corrections. In Fig.~\ref{fig:fpt_survival}~\textbf{C}, one can see that, on average, the plain activity has a similar effect as being initially oriented towards the wall, but at a way smaller scale, as it also encompasses particles departing away from it. Our results show that the survival probability decays as $S(t)\sim t^{-1/2}$, similar to the Brownian case. 
To understand the final decay, we leverage the final value theorem and examine the contribution of each correction individually:
\begin{subequations}
    \begin{equation}
        s\widehat{S}_{0}(s) \underset{s \to 0}{\sim}  \sqrt{s}Z_{0}, \label{eq:decay_s0}
    \end{equation}
    \begin{align}
        s\widehat{S}_{1}(s) &\underset{s \to 0}{\sim} \sqrt{s}\left( 1 - e^{Z_{0} \sqrt{\gamma}}\right) \frac{\cos(\vartheta_{0})}{\gamma}, \label{eq:decay_s1}\\
        s\widehat{S}_{2}(s) &\underset{s \to 0}{\sim}  \sqrt{s} \Big[ -3e^{2Z_{0} \sqrt{\gamma}} \left( Z_{0} \sqrt{\gamma} -2\right) \notag \\
        &\qquad -2e^{Z_{0} \sqrt{\gamma}} \big(3 + \cos(2\vartheta_{0}) \big) \Big] \frac{e^{-2Z_{0} \sqrt{\gamma}}}{12\gamma^{3/2}}. \label{eq:decay_s2}
    \end{align}
\end{subequations}
Equation~\eqref{eq:decay_s0} resolves the Brownian particles reaching at long times through diffusion ($\sqrt{s/D}$ in dimensional units), where the factor $\sqrt{s}$  leads to the $t^{-1/2}$ behavior. Despite the fact that the first- and second-order [Eqs. \eqref{eq:decay_s1}-\eqref{eq:decay_s2}] share the same scaling, they both present a non-trivial amplitude that depends on the initial orientation. Since the tail is scale-free, the relevance of the initial angle does not decay exponentially fast and the agent mostly remembers in which direction it was originally moving.



To study the interplay of the initial orientation and initial position, we compute the FPT probability density as a function of the initial angle~$\vartheta_{0}$ for different initial distances~$z_0$, which we compare to the particle's persistence length $l_{p}=v/D_{\mathrm{rot}} = a \mathrm{Pe}/\gamma$ [Fig.~\ref{fig:fpt_angle}]. The latter corresponds to the length the agent moves persistently before reorienting due to rotational diffusion. For a  particle that is initially at a small distance to the wall ($z_{0}/l_p =0.625$), the FPT density peaks at short times $ 10^{-2}\lesssim t/\tau \lesssim 10^{-1}$, irrespective of the initial orientation. Nevertheless, a more pronounced hue is observed for agents initially reaching towards the boundary ($\vartheta_{0}= \pi$). This anisotropy becomes striking when the particle is initially positioned at an intermediate distance ($z_{0}/l_p = 3.75$). Clearly, the FPT probability density is heavily dependent on the initial angle and is minimal when departing away from the boundary ($\vartheta_{0} = 0$). Lastly, if the agent is very far from the boundary ($z_{0}/l_p =25$), the memory of its initial orientation becomes less important and the FPT probability density does not exhibit any preferred direction.

\subsection{Contributions of different orders of the perturbation}\label{sec:results_FPTorders} 
\begin{figure}[tp]
\includegraphics[width = \linewidth]
{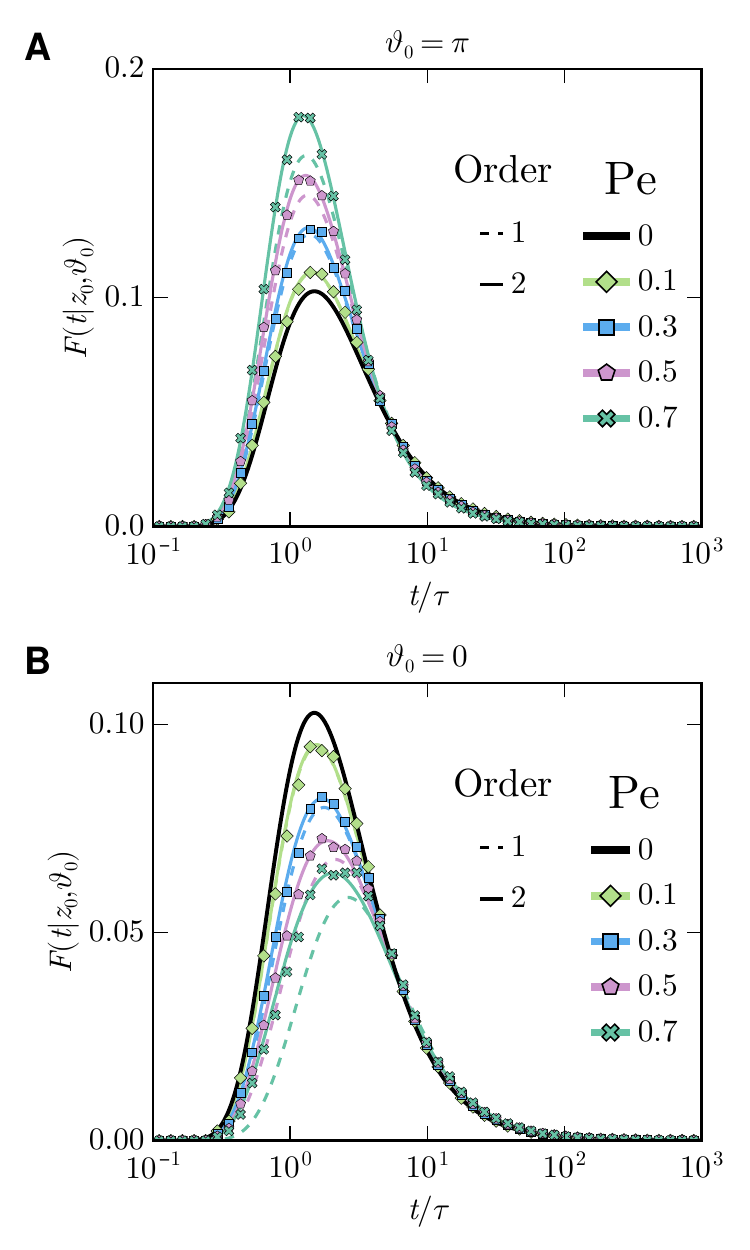}
\caption{Comparison of the FPT probability densities for two perturbation orders as a function of time $t$ for different P{\'e}clet numbers $\mathrm{Pe}$. The particle is initially \textbf{A}~facing the wall and \textbf{B}~ facing against the wall. The initial position is $z_{0}/a =3$. Symbols denote results obtained from simulations. \label{fig:error_fpt}}
\end{figure}

We note that including perturbations up to second-order provides good agreement between our theory and simulations. To understand the role of each correction, we compare them in Fig.~\ref{fig:error_fpt}. In the case of an agent initially oriented towards the wall (panel \textbf{A}), the first-order correction cannot predict the peak-value but the qualitative behavior is fair. However, when the agent departs away from the wall (panel \textbf{B}), the first-order correction performs worse. This is due to the more intricate dynamics in this case, since the agent can eventually reorient and reach the wall thanks to rotational diffusion. Mathematically, the simple $\cos(\vartheta_{0})$ in the first-order correction is not enough to describe such an activity-driven process, while it appears more suited for an agent initially moving towards the wall. The second-order perturbation corrects this effect and significantly improves the agreement between analytics and simulations. Furthermore, it includes constant ``purely active'' terms that do not depend on the initial orientation~$\vartheta_{0}$, and thus can capture the FPT statistics for randomly oriented agents [Fig.~\ref{fig:fpt_survival}~{\bf C}].

In principle, our expansion can be extended to arbitrary order. Here, we restricted the discussion to the second order. For the Péclet numbers discussed ($\mathrm{Pe} \lesssim 1$) the higher-order terms essentially do not contribute. On the other hand, we anticipate our expansion to perform worse for large P{\'e}clet numbers. It remains unclear if there is a finite radius of convergence for the expansion in $\mathrm{Pe}$ or if one simply needs to include higher-order terms to describe the regime of large Pe numbers.

\begin{figure*}[htp]
\includegraphics[width = \linewidth]{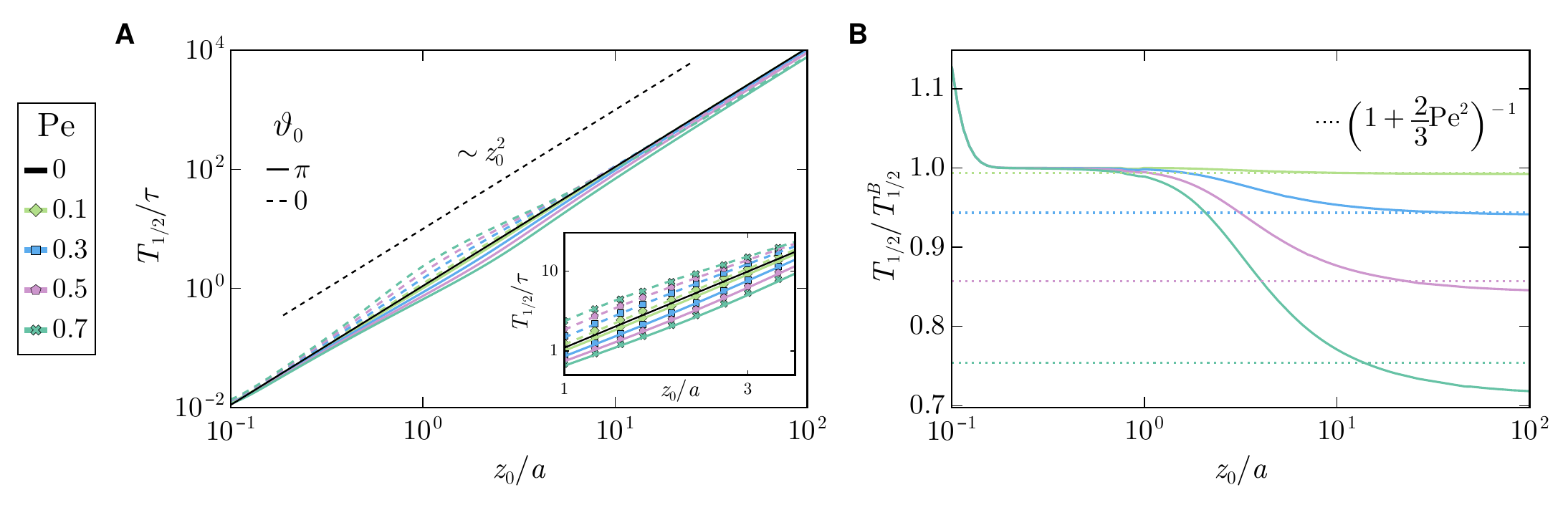}
\caption{ \textbf{A} Median of the FPT distribution $T_{1/2}$ as a function of the initial distance $z_{0}$ for different P{\'e}clet numbers $\mathrm{Pe}$ and two initial angles $\vartheta_{0}$. Numerical results are shown only in the inset for clarity. \textbf{B} Ratio of the median of an ABP and of a Brownian particle as a function of the initial distance $z_{0}$ for different P{\'e}clet numbers $\mathrm{Pe}$ and for an initial angle drawn from a uniform distribution $\mathcal{U}[0, 2\pi]$. The dotted line indicates the prediction $\left(1 + 2\mathrm{Pe}^{2}/3 \right)^{-1}$ for large initial distances~$z_0$.}
\label{fig:median_anisotropy}
\end{figure*}

\subsection{Median of the first-passage time \label{sec:results_median}}
Although the FPT distribution and survival probability yield interpretable results, they do not provide a quantitative answer to the question of which process is, on average, the fastest. As mentioned earlier, the mean FPT is a divergent quantity for the Brownian case and there is, in fact, no reason to expect the addition of low activity to change this. If $F_{t}$ denotes the random variable associated with the FPT, one can obtain its moments $\mathbb{E}[ F_{t}^{n}]$ through the relation:

\begin{align}
    \mathbb{E}[ F_{t}^{n}] &= (-1)^{n+1} \frac{\diff^{n}}{\diff s^{n}} \left[s \widehat{S}(s) \right]_{s=0},
\end{align}
which for the mean FPT reduces to $\mathbb{E} [F_{t}] = \widehat{S}(s=0)$. Our perturbation expansion does not change the diverging unperturbed term [Eq.\eqref{theory:perturb_s0}] and as such the mean FPT still diverges. In particular, in our case, agents that reach the wall after very long times, whether because they were initially departing from the wall and managed to reorient at later times due to rotational diffusion, or whether because they drifted away from it, count as outliers and will contribute to a broader tail (positive kurtosis) of the distribution and, hence, a diverging mean FPT. Alternatively, other quantities, such as the median~$T_{1/2}$, can give 
intelligible information about the completion of a  process. The latter is defined by
\begin{align}
    \int_{0}^{T_{1/2}}\diff t F(t |z_{0}, \vartheta_{0}) = \frac{1}{2},
\end{align}
representing, pathological cases aside, a distribution-unique quantity that splits the values into two, or the ``middle value''. Even if the median can greatly differ from the mean, it provides a non-outlier-skewed measure of the middle.

In Fig.~\ref{fig:median_anisotropy}~\textbf{A}, we show the behavior of the median for several P{\'e}clet numbers and for the situations of agents initially facing to and away from the wall. As reference case, we note that the median FPT of a Brownian particle ($\mathrm{Pe}=0$) can be computed analytically and assumes the form
\begin{equation}
    T_{1/2}^{B} = \frac{z_{0}^{2}}{4D \left[\text{erfc}^{-1}(1/2)\right]^2},\label{eq:median_brownian}
\end{equation}
where $\text{erfc}^{-1}$ denotes the  inverse complementary error function. Importantly, the median scales with~$T_{1/2}\propto z_0^2/D$, reflecting the time it takes to move a distance $z_0$ via diffusion (i.e. the diffusive time scale). As the activity increases, we observe at small $z_{0}$ a behavior close to that of a Brownian particle, indicating that diffusion dominates and takes the particle to the boundary. At intermediate initial distances, $z_{0}/a \sim 1$, the median departs from a ballistic regime and, depending on the initial orientation $\vartheta_{0}$, displays a smaller exponent ($\vartheta= \pi$) or larger exponent ($\vartheta= 0$). We notice that the case for a particle initially facing away from the wall, $\vartheta_{0}=0$ (dashed curve), ends up below the $\mathrm{Pe}=0$ curve for large~$z_{0} \gtrsim 10$. 
The two cases ($\vartheta_0= 0$ and $\vartheta_0=\pi$) eventually coalesce as $z_{0}$ increases further, as the relative importance of the initial orientation decreases conjointly. 

Considering that an ABP eventually reaches an effective diffusive regime at large times $t \gtrsim 1/D_{\mathrm{rot}}$ with effective diffusion coefficient (in 2D) \cite{Cates2013}
\begin{equation}
    D_{\mathrm{eff}} = D + \frac{v^{2}}{2D_{\mathrm{rot}}} = D \left( 1 + \frac{2}{3} \mathrm{Pe}^{2} \right),
\end{equation}
one can try to predict the median for large $z_0$ in the active case. Indeed, as the agent basically performs Brownian motion at large time and length scales, we expect to recover Eq.~\eqref{eq:median_brownian} with $D_{\mathrm{eff}}$ instead of $D$. This leads to the ratio: 
\begin{equation}
    \frac{T_{1/2}}{T_{1/2}^B} = \left(1 + \frac{2}{3}\mathrm{Pe}^{2} \right)^{-1}= 1- \frac{2}{3}{\mathrm{Pe}}^2+ O\left(\mathrm{Pe}^4\right), \label{eq:predic_ratio}
\end{equation}
indicating that active motion decreases the median at large $z_0$. 
The latter prediction is approached for large $z_0$ by the median corresponding to an averaged initial orientation, see Fig.~\ref{fig:median_anisotropy}~{\bf B}. We note that deviations become apparent for increasing P{\'e}clet numbers~${\mathrm{Pe}}\gtrsim0.7$, as the particle doesn't perfectly behave as a diffusing particle with diffuvisity $D_{\mathrm{eff}}$, but preserves the memory of its initial orientation [Eqs. \eqref{eq:decay_s1}-\eqref{eq:decay_s2}]. We further observe that at very small initial distances $z_{0}/a$, activity increases the median FPT, which can be ascribed to the other half of particles that initially move away from the boundary. 

Importantly, our results show that the median FPT of active agents (at intermediate and large initial distances) is smaller than for the passive case, irrespective of the P{\'e}clet number.

\subsection{Measure of the anisotropy \label{sec:results_anisotropy}}
To explore the effect of the initial orientation $\vartheta_0$ on the median, we introduce the anisotropy as the following function:
\begin{equation}
    \mathcal{A}( z_{0}, \vartheta_{0}) = \frac{T_{1/2}(z_{0}, \vartheta_{0} ) }{ T_{1/2} (z_{0}, \vartheta_{0} + \pi)}.
\end{equation}
It measures the ratio of the median of the FPT given an initial distance $z_{0}$ and initial orientation $\vartheta_{0}$ with the median given the same initial distance $z_{0}$ but with the diametrically opposed initial orientation $\vartheta_{0} + \pi$. 

\begin{figure}[tp]
\includegraphics[width = \linewidth]
{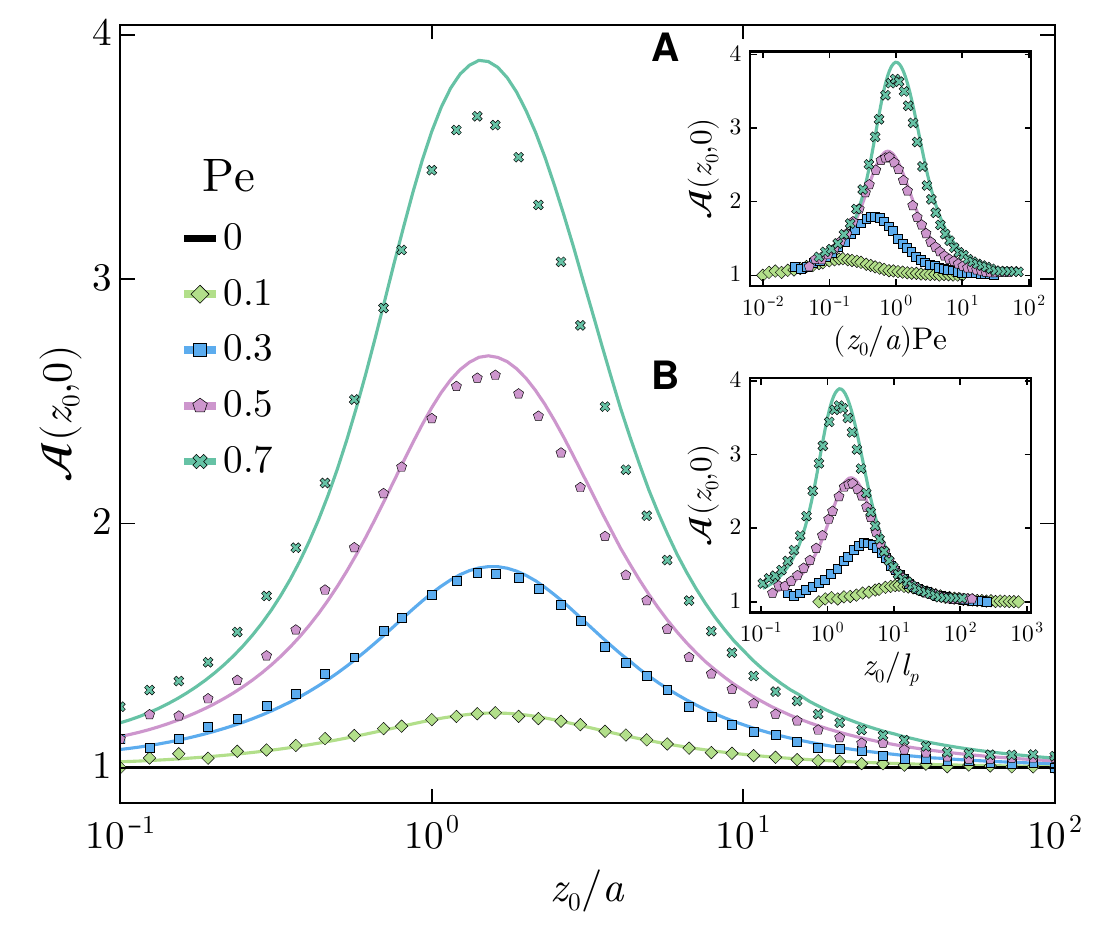}
\caption{Anisotropy $\mathcal{A}(z_{0},\vartheta_{0})$ as a function of the initial distance $z_{0}$ for different P{\'e}clet numbers $\mathrm{Pe}$. Lines represent theory and symbols correspond to simulation results. ({\it Inset \textbf{A}}) Anisotropy as a function of the rescaled initial distance  $(z_0/a) \mathrm{Pe}$.  ({\it Inset \textbf{B}}) Anisotropy as a function of the rescaled initial distance $z_0/l_{p}$, where $l_{p}$ denotes the persistence length. \label{fig:anisotropy}}
\end{figure}

 Fig.~\ref{fig:anisotropy} shows the anisotropy $\mathcal{A}(z_{0}, \vartheta_{0}=0)$ for various P{\'e}clet numbers as a function of the initial distance~$z_{0}$. It reaches its maximum at intermediate distances close to $z_{0}/a\sim 1.0$, which  shifts to smaller~$z_0$ and becomes more pronounced for increasing $\mathrm{Pe}$. We anticipate that the maximum should start to emerge at a distance $z_{0}$ where the contribution of persistent motion is comparable to translational diffusion. The length traveled via diffusion during time $\tau$ is $z_{0}^{2}\simeq \tau D$, while active motion leads to $z_{0}\simeq \tau v$. Taking these together, gives $z_{0}\simeq D/v \propto {\mathrm{Pe}}^{-1}$, which appears to confirm the shift towards shorter distances for increasing $\mathrm{Pe}$ and requires numerical confirmation for larger $\mathrm{Pe}$. This can be also be seen by rescaling with the P{\'e}clet number, which leads to a data collapse at small $z_{0}$ and a shift of the peak to $z_{0} \mathrm{Pe}\simeq 1$ for $\mathrm{Pe}=0.7$ [Fig.~\ref{fig:anisotropy}~{\textit{inset}~\textbf{A}}]. 

To gain further insights into the anisotropy, we also rescale the initial distance $z_{0}$ by the agent's persistence length~$l_{p}\propto {\mathrm{Pe}}$ [Fig.~\ref{fig:anisotropy}~\textit{inset}~\textbf{B}]. We observe that the curves at large  $z_{0}\gtrsim l_{p}$ collapse across all Pe, indicating that, at these distances (and at least for the particles that reach the wall) the initial orientation looses more and more relevance for their approach to the wall. 

\section{Summary and Conclusions}

Here, we have studied the FPT statistics of an active Brownian particle to reach an absorbing wall in two dimensions. Our perturbation approach, valid in the low-activity regime, allows computing exact expressions for the survival probability and first-passage time distribution. In contrast to bare diffusion, the FPT statistics crucially depend on the initial orientation of the active agent relative to the absorbing boundary. While the mean FPT remains divergent, the median can capture the effect of activity: it becomes reduced for non-zero P{\'e}clet numbers and large initial distances, while it displays more complex behavior at small and intermediate initial distances. To further quantify this effect, we have introduced an anisotropy measure, which displays a prominent peak at intermediate initial distances. 

We have here attacked the problem for low P{\'e}clet numbers and therefore the question naturally arises as to what happens when the activity fully dominates. Studying the median, we have seen a deviation from the quadratic $z_{0}^{2}$ regime at intermediate initial distances that becomes more prominent as $\mathrm{Pe}$ increases. If activity dominates, the agent essentially moves straight, and it would be interesting to see if the median follows a simple power-law dictated by the active time scale $z_0/v$. In addition, more complex dynamics, including circular~\cite{kummelCircularMotionAsymmetric2013,kurzthalerIntermediateScatteringFunction2017} or run-reverse motion~\cite{taktikosHowMotilityPattern2013}, are expected to amplify these predictions, thus opening new research directions.

While the ABP model serves as minimal coarse-grained model for accurately describing the dynamics of active agents in a bulk environment, the presence of real boundaries can play a major role through, for example, long-ranged hydrodynamic interactions~\cite{daddi-moussa-iderHydrodynamicsCanDetermine2021,piroOptimalNavigationMicroswimmers2022}.
In effect, the finite distance to the wall naturally introduces a backflow that needs to be taken into account~\cite{ spagnolieHydrodynamicsSelfpropulsionBoundary2012b, spagnolieGeometricCaptureEscape2015} and affects the dynamics of active agents differently, depending on the flow signature produced by their swimming mechanism. These effects have been thoroughly studied in the context of surface accumulation~\cite{berkeHydrodynamicAttractionSwimming2008,liAccumulationMicroswimmersSurface2009,drescherBiofilmStreamersCause2013, schaarDetentionTimesMicroswimmers2015} and near-surface motion~\cite{laugaSwimmingCirclesMotion2006}, yet the effect of hydrodynamics on the FPT statistics still remains to be elucidated.

When studying FPT statistics, another major impact of hydrodynamics concerns the effects on the overall stochasticity through the noise at play. Indeed, it is well-established that the diffusion coefficients depend heavily on the distance to the wall~\cite{bian111YearsBrownian2016,alexandreNonGaussianDiffusionSurfaces2023a}. This is expected to be even more pronounced in the case of a non-flat wall, where wall roughness and surface topography can induce changes in the agent's dynamics via modifying its hydrodynamic mobility~\cite{huangEffectInterfacesNearby2015,kurzthalerParticleMotionNearby2020, chaseHydrodynamicallyInducedHelical2022}. 

\begin{acknowledgments} 
Y.B. gratefully acknowledges Vincenzo M. Schimmenti and Paolo Pigato for insightful comments. T.F. and M.L.G. acknowledge funding by the Austrian Science Fund (FWF): P~35580-N ‘Target Search of Single Active Brownian Particles and Run-and-Tumble Agents’.
\end{acknowledgments}

\begin{widetext}
\section*{Appendix}
\appendix
The appendix contains additional analytical results of our perturbation theory [Sec.~\ref{appendix:analytics}] and details of our simulation set-up [Sec.~\ref{appendix:simulations}].

\section{Perturbation theory}
\subsection{Additional analytical results \label{appendix:analytics}}

In this section, we give the expression of the second-order correction to the propagator $\widehat{\mathbb{P}}_{2}$. For $Z \leq Z_{0}$, we have
\begin{subequations}
\begin{align}
    &\widehat{\mathbb{P}}_{2}^{-}(s,Z | Z_{0}, \vartheta_{0}) = -\frac{1}{48 \gamma^{2}} \Bigg( \frac{6e^{-\left(Z+Z_{0} \right)\left(\sqrt{s}+\sqrt{s+\gamma} \right) }}{\sqrt{s} \left( \sqrt{s} + \sqrt{s+\gamma} \right)  } \Bigg[e^{2 \sqrt{s}Z +\left(Z+Z_{0} \right) \sqrt{s+\gamma}} \left( 2s + \gamma + \sqrt{s}(Z-Z_{0})\gamma  \right) \left( \sqrt{s} + \sqrt{s+\gamma}\right) \notag \\ 
    &-2e^{\sqrt{s} \left( Z+Z_{0}\right)} \sqrt{s} \left( s+\gamma + \sqrt{s(s+\gamma)} \right) + 4e^{\sqrt{s} Z_{0} + Z \sqrt{s+\gamma}}\sqrt{s} \left( s+\gamma + \sqrt{s (s+\gamma)} \right) \notag\\
    &- 2e^{\sqrt{s}(Z+Z_{0}) + 2Z \sqrt{s+\gamma}}\sqrt{s} \left( s+ \gamma + \sqrt{s(s+\gamma)} \right)  
    +4e^{\sqrt{s}Z + Z_{0} \sqrt{s+\gamma}} \sqrt{s} \left( s+\gamma + \sqrt{s(s+\gamma)}\right) \notag  \\
    &+e^{\left( Z + Z_{0}\right) \sqrt{s+\gamma} } \left(-6s^{3/2} + s \left(Z +Z_{0} \right) \gamma -6s \sqrt{s+\gamma} -\gamma \sqrt{s+\gamma} + \sqrt{s} \gamma \left( -5 + \left( Z+Z_{0}\right) \sqrt{s+\gamma} \right) \right) \Bigg] \notag \\
    &+\frac{\cos(2\vartheta_{0})}{\sqrt{s+\gamma}} \Bigg[4e^{ \left(Z-Z_{0} \right) \sqrt{s+\gamma}} \left( s+\gamma \right) + 4e^{-\left( Z+Z_{0}\right) \sqrt{s+\gamma}} \left( s+\gamma \right) -8 e^{-\sqrt{s}Z - Z_{0} \sqrt{s+\gamma}}(s+\gamma)\notag\\ &+ 8e^{-\sqrt{s}Z - Z_{0} \sqrt{s+4\gamma}} \left( s+\gamma \right) 
    -8e^{-Z \sqrt{s+\gamma} - Z_{0}\sqrt{s+4\gamma}} \left( s+\gamma \right) - 3e^{\sqrt{s} \left( Z-Z_{0}\right)}\sqrt{s \left(s +\gamma \right)}\notag\\ &+ 3e^{-\sqrt{s} \left( Z + Z_{0} \right)} \sqrt{s \left( s+\gamma \right)} - e^{\left(Z-Z_{0} \right) \sqrt{s+4\gamma}} \sqrt{\left( s+\gamma \right) \left( s+4\gamma \right)} + e^{- \left( Z+Z_{0} \right) \sqrt{s+4\gamma}} \sqrt{\left( s+\gamma \right) \left( s+4\gamma \right)}
    \Bigg]\Bigg),\label{eq:p2m}
\end{align}
and for $Z \geq Z_{0}$, we obtain
\begin{align}
    &\widehat{\mathbb{P}}_{2}^{+}(s,Z | Z_{0}, \vartheta_{0}) = - \frac{1}{48 \gamma^{2} \sqrt{s \left( s+\gamma \right)}} \Bigg( 
    3\gamma\sqrt{s+\gamma}\Bigg[ \frac{1}{\gamma} \Big( e^{\sqrt{s} \left( -3Z +Z_{0} \right)}\gamma - e^{-\sqrt{s}\left(3Z+Z_{0} \right)} \gamma -4e^{\left(-Z + Z_{0} \right) \sqrt{s+\gamma}} \sqrt{s \left( s + \gamma \right)} \notag \\
    &-4e^{- \left( Z + Z_{0} \right) \sqrt{s+\gamma}} \sqrt{s \left( s + \gamma \right)} + 8e^{-\sqrt{s}Z_{0} -Z \sqrt{s+\gamma}} \sqrt{s \left( s + \gamma \right)} + 8e^{-\sqrt{s}Z - Z_{0} \sqrt{s+\gamma}}\sqrt{s \left( s + \gamma \right)} \notag \\
    &+ 2e^{\sqrt{s} \left( -Z + Z_{0} \right)} \left( 2s + \gamma + \sqrt{s} \left( -Z + Z_{0} \right) \gamma \right) - 2e^{-\sqrt{s} \left( Z + Z_{0} \right)} \left( 2s + \gamma - \sqrt{s} \left( Z+ Z_{0} \right) \gamma +4 \sqrt{s \left( s+ \gamma \right)} \right)\Big) \notag  \\
    &- 2e^{-3\sqrt{s}Z} \sinh(\sqrt{s} Z_{0})
    \Bigg] \notag \\ 
    &+ \cos(2\vartheta_{0}) \Bigg[- \frac{2e^{-2\sqrt{s}Z - \left( Z+ Z_{0} \right)\left( \sqrt{s+\gamma} + \sqrt{s+4\gamma} \right) }}{\left( \sqrt{s} + \sqrt{s+\gamma}\right) \left( \sqrt{s} + \sqrt{s+4\gamma} \right)} \left( -1 + e^{2\sqrt{s}Z} \right) \gamma \sqrt{s+\gamma}  \Big( e^{\left( Z+ Z_{0}\right) \sqrt{s+\gamma}} \sqrt{s+4\gamma} \left( \sqrt{s} + \sqrt{s+\gamma} \right)  \notag \\
    & - e^{Z \sqrt{s+\gamma} + Z_{0} \sqrt{s+\gamma} + 2Z_{0} \sqrt{s+4\gamma}} \sqrt{s+4\gamma} \left( \sqrt{s} + \sqrt{s+\gamma} \right) + e^{\left( Z + Z_{0} \right) \sqrt{s+4\gamma}} \sqrt{s+\gamma} \left( \sqrt{s} + \sqrt{s+4\gamma} \right) \notag \\
    &-2e^{Z_{0} \sqrt{s+\gamma} + Z \sqrt{s+4\gamma}} \sqrt{s+\gamma} \left( \sqrt{s} + \sqrt{s+4\gamma} \right)  + e^{Z \sqrt{s+4\gamma} + Z_{0} \left( 2\sqrt{s+\gamma} + \sqrt{s+4\gamma}\right)} \sqrt{s+\gamma} \left( \sqrt{s} + \sqrt{s+4\gamma} \right) \Big) \notag \\
    & + e^{-\sqrt{s}Z -2Z_{0} \left( \sqrt{s+\gamma} + \sqrt{s+4\gamma} \right)} \Big( 8 \left( e^{Z_{0} \left( 2\sqrt{s+\gamma} + \sqrt{s+4\gamma} \right)} - e^{Z_{0} \left( \sqrt{s+\gamma} + 2\sqrt{s+4\gamma} \right)} \right) \sqrt{s} \left( s+\gamma \right) \notag \\ 
    &+ \sqrt{s}\Big[4e^{2Z_{0}\sqrt{s+4\gamma}} \left( s +  \gamma \right) -8e^{Z_{0} \left( \sqrt{s+\gamma} + \sqrt{s + 4\gamma} \right)} \left( s+ \gamma \right) +e^{2Z_{0} \sqrt{s+\gamma} }\sqrt{\left( s+\gamma \right) \left( s+ 4\gamma \right)} \notag \\ 
    &+ e^{2Z_{0} \left( \sqrt{s+\gamma} + \sqrt{s+4\gamma} \right)} \left( 4s + 4\gamma - \sqrt{\left( s+ \gamma \right) \left( s+ 4\gamma \right) } \right) \Big] \cosh(\sqrt{s} Z_{0}) + \sqrt{s+\gamma} \Big[ -3e^{2Z_{0} \left( \sqrt{s+\gamma} + \sqrt{s+4\gamma} \right)}s\notag \\ 
    &+ 4e^{2Z_{0} \sqrt{s+4\gamma}} \left( s + \gamma \right) 
    -8e^{Z_{0} \left( \sqrt{s+\gamma} + \sqrt{s+4\gamma} \right) } \left( s + \gamma \right) + e^{2Z_{0} \sqrt{s+\gamma}} \left( s+ 4\gamma \right) \Big] \sinh(\sqrt{s}Z_{0})\Big) \notag \\
    &- 2\gamma \sqrt{s+\gamma}  \Big( e^{-Z_{0} \left( \sqrt{s} + \sqrt{s+\gamma} + 2\sqrt{s+4\gamma}\right)} \Big[ - \frac{e^{-\sqrt{s}Z - Z_{0} \sqrt{s+\gamma} + Z_{0} \sqrt{s+4\gamma} }}{\sqrt{s} + \sqrt{s+\gamma}} \Big( -2e^{Z_{0} \sqrt{s+\gamma}} + e^{Z_{0} \sqrt{s+4\gamma}} \notag\\
    &+ e^{Z_{0} \left( 2\sqrt{s+\gamma} + \sqrt{s+4\gamma} \right)} \Big) \sqrt{s+\gamma}  
    + \frac{e^{-\sqrt{s}Z + Z_{0} \sqrt{s+\gamma}} }{\sqrt{s} + \sqrt{s+4\gamma}}\left( -1 + e^{2Z_{0}\sqrt{s+4\gamma}} \right) \sqrt{s+4\gamma} \notag \\ 
    &- \frac{e^{-\sqrt{s} \left( Z-2Z_{0}  \right)}}{4\gamma} \Big( -4e^{Z_{0} \left( -\sqrt{s+\gamma} + \sqrt{s+4\gamma} \right)} \left( -2e^{Z_{0} \sqrt{s+\gamma}} + e^{Z_{0} \sqrt{s+4\gamma}} + e^{Z_{0} \left( 2\sqrt{s+\gamma} + \sqrt{s+4\gamma} \right)} \right) \left( s+\gamma + \sqrt{s \left( s+ \gamma \right)} \right)  \notag \\ 
    &+ e^{Z_{0}\sqrt{s+\gamma}}\left( -1 + e^{2Z_{0} \sqrt{s+4\gamma}} \right) \left(  s+ 4\gamma + \sqrt{s \left( s+4\gamma \right)}\right)   \Big) \Big] \notag \\
    &+  e^{-Z \left( 2\sqrt{s} + \sqrt{s+\gamma} + \sqrt{s+4\gamma} \right)} \Big[ \frac{e^{Z \sqrt{s+4\gamma} - Z_{0} \left( \sqrt{s+\gamma} + \sqrt{s+4\gamma} \right) }}{\sqrt{s} + \sqrt{s+\gamma}} \left(-2e^{Z_{0} \sqrt{s+\gamma}} + e^{Z_{0} \sqrt{s+4\gamma}} + e^{Z_{0} \left( 2\sqrt{s+\gamma} + \sqrt{s+4\gamma} \right) } \right) \sqrt{s+\gamma} \notag \\ 
    &+ \frac{e^{2\sqrt{s} Z - Z_{0} \sqrt{s+4\gamma}}}{4\gamma} \Big( -4e^{Z_{0} \sqrt{s+\gamma} + Z \sqrt{s+4\gamma}} \left( -2e^{Z_{0}\sqrt{s+\gamma}} + e^{Z_{0}\sqrt{s+4\gamma}} + e^{Z_{0} \left( 2\sqrt{s+\gamma} + \sqrt{s+4\gamma} \right)} \right)  \left( s+\gamma + \sqrt{s \left( s+ \gamma \right)} \right) \notag \\
    &+ e^{Z \sqrt{s+\gamma}} \left( -1 + e^{2Z_{0} \sqrt{s+4\gamma} }\right) \left( s+ 4\gamma + \sqrt{s \left( s + 4\gamma\right) } \right) \Big) - \frac{2e^{Z \sqrt{s+\gamma}} \sqrt{s+4\gamma} }{\sqrt{s} + \sqrt{s+4\gamma}} \sinh(Z_{0} \sqrt{s+4\gamma} )
    \Big] \Big) \Bigg]\Bigg). \label{eq:p2p}
\end{align}
\end{subequations}
\subsection{Recursive scheme for higher-order perturbations \label{appendix:recursive_scheme}}
Here, we present a scheme to compute the $n-$th order correction to the propagator. Therefore, we first define 
\begin{align}
    K^{(0)}_{(m)}(Z,Z_0) &= \frac{1}{2p_{m}} \left( e^{-p_{m}|Z-Z_{0}|} -  e^{-p_{m}|Z+Z_{0}|} \right),    \label{kernel_green}  
\end{align}
as the kernel of order zero. The kernel of order one is then defined as:
\begin{align}
    K_{(l, m)}^{(1)}(Z,Z_0) &= \int_{0}^{\infty}\mathrm{d}Y~  K^{(0)}_{(l)}(Y, Z_{0}) \left(  -\frac{\partial}{\partial_{Z}}K_{(m)}^{(0)}(Z,Z_{0}) \biggr\rvert_{Z=Y}\right).  \label{eq:klm^1}
\end{align}
In the same way, one can recursively define a kernel of order $j$ by:
\begin{align}
    K_{(l, m, ...)}^{(j)}(Z,Z_0) &= \int_{0}^{\infty}\mathrm{d}Y~  K^{(0)}_{(l)}(Y, Z_{0}) \left(  -\frac{\partial}{\partial_{Z}}K_{(m, ...)}^{(j-1)}(Z,Z_{0}) \biggr\rvert_{Z=Y}\right). 
\end{align}

\begin{figure*}[tp]
\includegraphics[width = \linewidth]{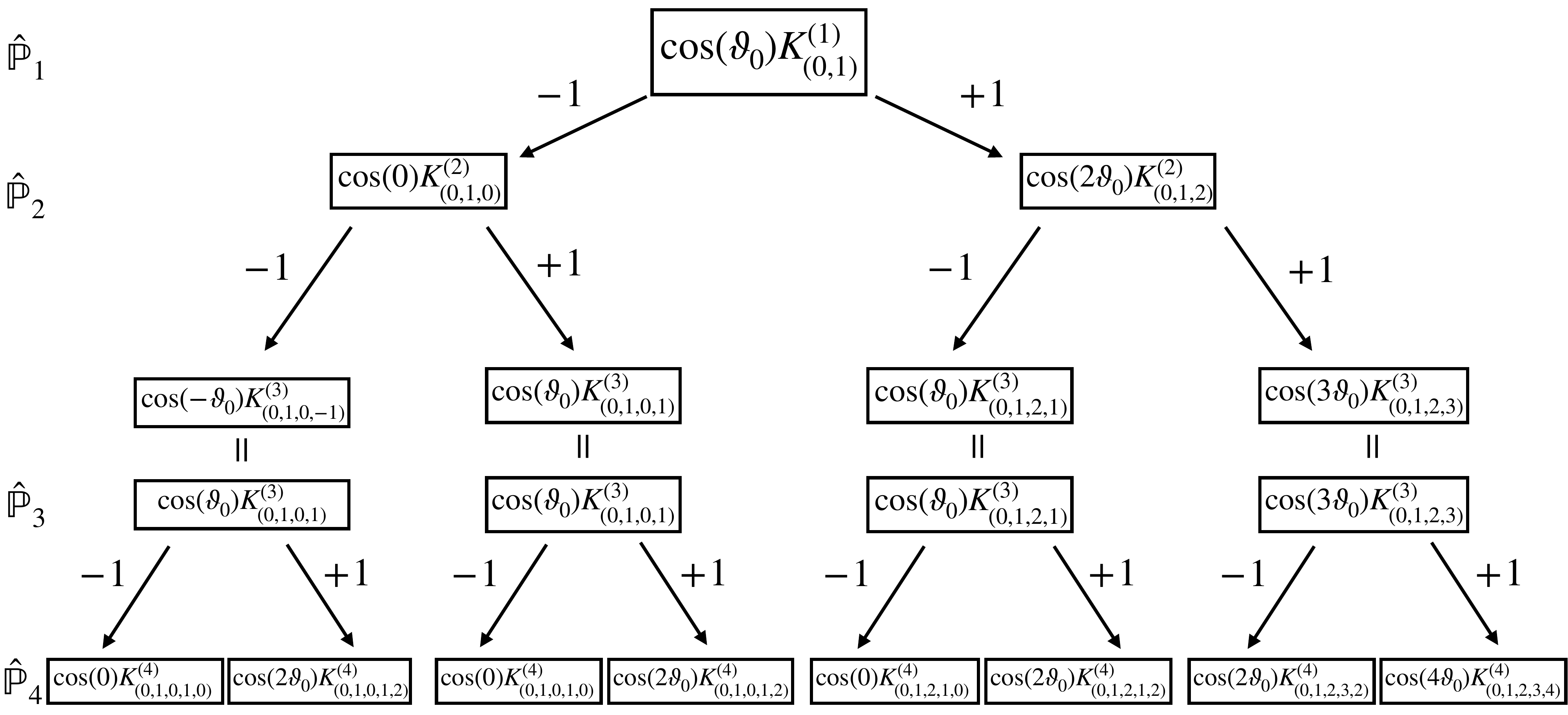}
\caption{Scheme to iteratively obtain the correction of the next order: at each step, the argument of the cosine is decreased and increased by $\vartheta_{0}$ and the kernel's order is increased. We use the symmetry $K_{(-\ell)} = K_{(\ell)}$ for the subindices and the fact that the cosine is an even function.}
\label{fig:appendix_scheme}
\end{figure*}

Then, using the previous notation and Eq. \eqref{eq:pi} in the main text, the first-order perturbation to the propagator is
\begin{align}
    \widehat{\mathbb{P}}_{1}(s,Z | Z_{0}, \vartheta_{0}) =  \cos(\vartheta_{0}) K_{(0,1)}^{(1)}(Z, Z_{0}).
    \label{eq:appendix_perturb1}
\end{align}

The next-order perturbation is obtained by increasing the kernel's order by one, as well as adding and subtracting $\pm1$ to the kernel's sub-indices and the multiple of $\vartheta_{0}$ in the cosine's argument (see Fig.~\ref{fig:appendix_scheme}). The final expression is then divided by $2^{n-1}$, where $n$ is the perturbation's order. For instance, the second-order perturbation to the propagator is given by:
    \begin{align}
    \begin{split}
    &\widehat{\mathbb{P}}_{2}(s,Z,| Z_{0}, \vartheta_{0}) =  \frac{1}{2}  \Bigg[ K^{(2)}_{(0,1,0)}(Z,Z_{0}) + \cos(2 \vartheta_{0})  K^{(2)}_{(0,1,2)}(Z,Z_{0})  \Bigg],
    \end{split}
    \end{align}
    which can be readily checked with Eqs.~\eqref{eq:p2m}-\eqref{eq:p2p}. Higher-order perturbations can be iteratively derived. For example, the third and fourth order can be computed via
    \begin{subequations}
    \begin{align}
    &\widehat{\mathbb{P}}_{3}(s,Z,| Z_{0}, \vartheta_{0}) =  \frac{1}{4} \Bigg[ \cos(3\vartheta_{0}) K^{(3)}_{(0,1,2,3)}(Z,Z_{0}) + \cos(\vartheta_{0}) K^{(3)}_{(0,1,2,1)}(Z,Z_{0}) + 2\cos(\vartheta_{0}) K^{(3)}_{(0,1,0,1)}(Z,Z_{0})  \Bigg], \\ 
    &\widehat{\mathbb{P}}_{4}(s,Z,| Z_{0}, \vartheta_{0}) =  \frac{1}{8} \Bigg[ \cos(4\vartheta_{0}) K^{(4)}_{(0,1,2,3,4)}(Z,Z_{0}) + 2\cos(2\vartheta_{0}) K^{(4)}_{(0,1,0,1,2)}(Z,Z_{0}) + \cos(2\vartheta_{0}) K^{(4)}_{(0,1,2,1,2)}(Z,Z_{0})\notag \\ 
    &\qquad \qquad \qquad \qquad \qquad  + 2K^{(4)}_{(0,1,0,1,0)}(Z,Z_{0})  + K^{(4)}_{(0,1,2,1,0)}(Z,Z_{0}) + \cos(2\vartheta_{0}) K^{(4)}_{(0,1,2,3,2)}(Z,Z_{0}) \Bigg]. 
    \end{align}
    \end{subequations}
\section{Computer Simulations \label{appendix:simulations}}
To perform stochastic simulations, we discretize Eqs.~\eqref{eq:stoch_r} and \eqref{eq:stoch_theta} according to the Euler-Maruyama scheme:
\begin{subequations}
\begin{align}
\vec{r}(t+ \Delta t) &=  \vec{r}(t) + v \vec{e}(\vartheta(t)) \Delta t + \sqrt{2D \Delta t} \vec{N}_{t}(0, 1), \\ 
\vartheta(t+ \Delta t) &= \vartheta(t) +\sqrt{2D_{\mathrm{rot}} \Delta t} N_{r}(0, 1),
\end{align}
\end{subequations}
\noindent where $\Delta t = 10^{-3} \tau$ is the time step and $\vec{N}_{t}(0, 1)$ and $N_{r}(0, 1)$ are independent, normally distributed random variable with zero mean and unit variance. Furthermore, the statistics are obtained by simulating trajectories of $10^5$ particles.
\end{widetext}

\bibliography{bibliography_abp}

\end{document}